\documentclass[
    ,final            % use final for the camera ready runs
  ]
  {aipproc}

\layoutstyle{6x9}

\newcommand{\text}[1]{{#1}}

\newcommand{\ba}{\begin{eqnarray}}
\newcommand{\ea}{\end{eqnarray}}

\begin{document}

\title{Pentaquark spectroscopy: exotic $\Theta$ baryons}

\author{R. Bijker}
{address={Instituto de Ciencias Nucleares, 
Universidad Nacional Aut\'onoma de M\'exico, 
A.P. 70-543, 04510 M\'exico, D.F., M\'exico}}

\author{M.M. Giannini}
{address={Dipartimento di Fisica dell'Universit\`a di Genova, 
I.N.F.N., Sezione di Genova, via Dodecaneso 33, 16164 Genova, Italy}}

\author{E. Santopinto}
{address={Dipartimento di Fisica dell'Universit\`a di Genova, 
I.N.F.N., Sezione di Genova, via Dodecaneso 33, 16164 Genova, Italy}}

\begin{abstract}
We propose a collective stringlike model of $q^4\bar{q}$ pentaquarks with 
the geometry of an equilateral tetrahedron in which the four quarks are 
located at the four corners and the antiquark in its center. The nonplanar 
equilibrium configuration is a consequence of the permutation symmetry of 
the four quarks. In an application to the spectrum of exotic $\Theta$ baryons, 
we find that the ground state pentaquark has angular momentum and parity 
$J^p=1/2^-$ and a small magnetic moment of 0.382 $\mu_N$. The decay width 
is suppressed by the spatial overlap with the decay products. 
\end{abstract}

\maketitle

\section{Introduction}

The building blocks of atomic nuclei, the nucleons, are composite extended 
objects, as is evident from their anomalous magnetic moment, their excitation 
spectrum and their charge distribution. Examples of excited states of the 
nucleon are the $\Delta$(1232) and N(1440) Roper resonances. To first 
approximation, the internal structure of the nucleon at low energy 
can be ascribed to three bound constituent quarks $q^3$.  

The discovery of the $\Theta(1540)$ with positive strangeness $S=+1$ by 
the LEPS Collaboration \cite{leps} as the first example of an exotic baryon 
(with quantum numbers that can not be obtained with $q^3$ configurations) 
has motivated an enormous amount of experimental and 
theoretical studies. The width of this state is observed to be very small 
$< 20$ MeV (or perhaps as small as a few MeV's). More recently, the NA49 
Collaboration \cite{cern} reported evidence for the existence of another 
exotic baryon $\Xi(1862)$ with strangeness $S=-2$. The $\Theta^+$ and 
$\Xi^{--}$ resonances are interpreted as $q^4 \bar{q}$ pentaquarks belonging 
to a flavor antidecuplet with quark structure $uudd\bar{s}$ and $ddss\bar{u}$, 
respectively. In addition, there is now the first evidence \cite{h1} for a 
heavy pentaquark $\Theta_c(3099)$ in which the antistrange quark in the 
$\Theta^+$ is replaced by an anticharm quark. The spin and parity of these 
states have not yet been determined experimentally. 
For a review of the experimental status we refer to \cite{zhaoclose}. 

Theoretical interpretations range from chiral soliton models \cite{soliton} 
which provided the motivation for the experimental searches, correlated 
quark (or cluster) models \cite{cluster}, and various constituent quark 
models \cite{cqm,BGS}. A review of the theoretical literature of 
pentaquark models can be found in \cite{jennings}. 

In this contribution, we introduce a collective stringlike model of 
$q^4\bar{q}$ pentaquarks in which the four quarks are located at the four 
corners of an equilateral tetrahedron and the antiquark in its center. 
This nonplanar equilibrium configuration is a consequence of the permutation 
symmetry of the four quarks \cite{BGS1}. As an application, we discuss the 
spectrum of exotic $\Theta$ baryons, as well as the parity and magnetic 
moments of the ground state decuplet baryons. The decay width is suppressed 
by the spatial overlap with the decay products. 

\section{Pentaquark states}

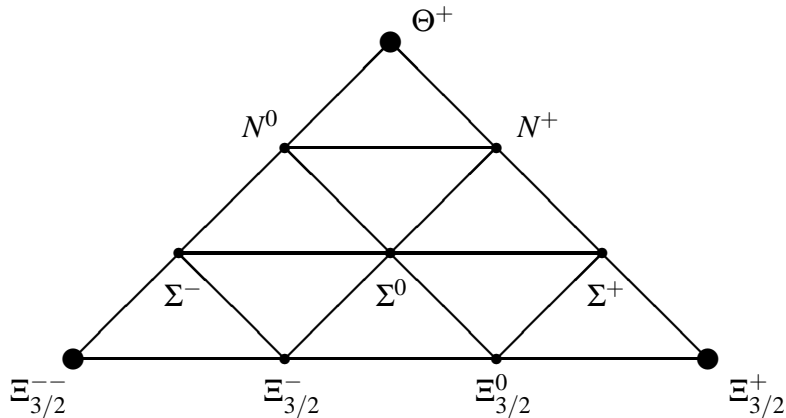
\begin{figure}[b]
\centering
\setlength{\unitlength}{0.8pt}
\begin{picture}(350,185)(70,80)
\thicklines
\put(200,200) {\line(1,0){100}}
\put(150,150) {\line(1,0){200}}
\put(100,100) {\line(1,0){300}}
\put(100,100) {\line(1,1){150}}
\put(200,100) {\line(1,1){100}}
\put(300,100) {\line(1,1){ 50}}
\put(200,100) {\line(-1,1){ 50}}
\put(300,100) {\line(-1,1){100}}
\put(400,100) {\line(-1,1){150}}
\multiput(250,250)(100,0){1}{\circle*{5}}
\multiput(200,200)(100,0){2}{\circle*{5}}
\multiput(150,150)(100,0){3}{\circle*{5}}
\multiput(100,100)(100,0){4}{\circle*{5}}
\multiput(100,100)(300,0){2}{\circle*{10}}
\put(250,250){\circle*{10}}
\put(260,255){$\Theta^+$}
\put(180,205){$N^0$}
\put(310,205){$N^+$}
\put(143,125){$\Sigma^-$}
\put(243,125){$\Sigma^0$}
\put(343,125){$\Sigma^+$}
\put( 70, 80){$\Xi_{3/2}^{--}$}
\put(190, 80){$\Xi_{3/2}^{-}$}
\put(290, 80){$\Xi_{3/2}^{0}$}
\put(410, 80){$\Xi_{3/2}^+$}
\end{picture}
\caption{$SU(3)$ antidecuplet. The isospin-hypercharge multiplets are 
$(I,Y)=(0,2)$, $(\frac{1}{2},1)$, $(1,0)$ and $(\frac{3}{2},-1)$. 
Exotic states are indicated with $\bullet$.}
\label{flavor33e}
\end{figure}

We consider pentaquarks to be built of five constituent parts whose 
internal degrees of freedom are taken to be the three light flavors 
$u$, $d$, $s$ with spin $s=1/2$ and three colors $r$, $g$, $b$. 
The corresponding algebraic structure consists of the usual spin-flavor 
and color algebras $SU_{\rm sf}(6) \otimes SU_{\rm c}(3)$. The full 
decomposition of the spin-flavor states into spin and flavor states can 
be found in \cite{BGS}
\ba
SU_{\rm sf}(6) \supset SU_{\rm f}(3) \otimes SU_{\rm s}(2) 
\supset SU_{\rm I}(2) \otimes U_{\rm Y}(1) \otimes SU_{\rm s}(2) ~.
\label{sfchain}
\ea
The allowed flavor multiplets are singlet, octet, decuplet, antidecuplet, 
27-plet and 35-plets. The first three have the same values of the isospin 
$I$ and hypercharge $Y$ as $q^3$ systems. However, the antidecuplets, 
the 27-plets and 35-plets contain exotic states which cannot be obtained 
from three-quark configurations. These states are more easily identified 
experimentally due to the uniqueness of their quantum numbers. 
The recently observed $\Theta^+$ and $\Xi^{--}$ resonances are 
interpreted as pentaquarks belonging to a flavor antidecuplet with 
isospin $I=0$ and $I=3/2$, respectively. In Fig.~\ref{flavor33e} 
the exotic states are indicated by a $\bullet$: the $\Theta^+$ is an 
isosinglet $I=0$ with hypercharge $Y=2$ (strangeness $S=+1$), and the 
cascade pentaquarks $\Xi_{3/2}$ have hypercharge $Y=-1$ (strangeness $S=-2$) 
and isospin $I=3/2$. 

In the construction of the classification scheme we are guided by two 
conditions: the pentaquark wave function should be a color singlet and it 
should be antisymmetric under any permutation of the four quarks \cite{BGS}. 
The permutation symmetry of the four-quark system is given by $S_4$ 
which is isomorphic to the tetrahedral group ${\cal T}_d$. We use 
the labels of the latter to classify the states by their symmetry 
character: symmetric $A_1$, antisymmetric $A_2$ or mixed symmetric 
$E$, $F_2$ or $F_1$. 

The relative motion of the five constituent parts is described in terms 
of the Jacobi coordinates 
\ba
\begin{array}{ll}
\vec{\rho}_1 \;=\; \frac{1}{\sqrt{2}} ( \vec{r}_1 - \vec{r}_2 ) ~, & 
\vec{\rho}_3 \;=\; \frac{1}{\sqrt{12}} 
( \vec{r}_1 + \vec{r}_2 + \vec{r}_3 - 3\vec{r}_4 ) ~, \\
\vec{\rho}_2 \;=\; \frac{1}{\sqrt{6}} 
( \vec{r}_1 + \vec{r}_2 - 2\vec{r}_3 ) ~, \hspace{1cm} &
\vec{\rho}_4 \;=\; \frac{1}{\sqrt{20}} 
( \vec{r}_1 + \vec{r}_2 + \vec{r}_3 + \vec{r}_4 - 4\vec{r}_5 ) ~, 
\end{array}
\label{jacobi}
\ea
where $\vec{r}_i$ ($i=1,..,4$) denote the coordinate of the $i$-th quark,  
and $\vec{r}_5$ that of the antiquark. The last Jacobi coordinate 
is symmetric under the interchange of the quark coordinates, 
and hence transforms as $A_1$ under ${\cal T}_d$ ($\sim S_4$), whereas 
the first three transform as three components of $F_2$.  

The total pentaquark wave function is the product of the spin, flavor, 
color and orbital wave functions  
$\psi=\psi^{\rm s} \psi^{\rm f} \psi^{\rm c} \psi^{\rm o}$~. 
Since the color part of the pentaquark wave function is a singlet 
and that of the antiquark an anti-triplet, the color wave function of 
the four-quark configuration is a triplet with $F_1$ symmetry.  
The total $q^4$ wave function is antisymmetric ($A_2$), hence the 
orbital-spin-flavor part has to have $F_2$ symmetry 
\ba
\psi \;=\; \left[ \psi^{\rm c}_{F_1} \times 
\psi^{\rm osf}_{F_2} \right]_{A_2} ~.
\label{wf}
\ea
Here the square brackets $[\cdots]$ denote the tensor coupling under the 
tetrahedral group ${\cal T}_d$. 

\section{Stringlike model}

In this section, we discuss a stringlike model for pentaquarks, which is a 
generalization of a collective stringlike model developed for $q^3$ baryons 
\cite{BIL}. We introduce 
a dipole boson with $L^p=1^-$ for each independent relative coordinate, and 
an auxiliary scalar boson with $L^p=0^+$, which leads to a compact 
spectrum-generating algebra of $U(13)$ for the radial excitations. 
As a consequence of the invariance of the interations under the permutation 
symmetry of the four quarks, the most favorable geometric configuration is an 
equilateral tetrahedron in which the four quarks are located at the four 
corners and the antiquark in its center \cite{BGS1}. 
This configuration is in agreement with that of \cite{song} 
in which arguments based on the flux-tube model were used to suggest a 
nonplanar structure for the $\Theta(1540)$ pentaquark to explain its narrow 
width. In the flux-tube model, the strong color field between a pair of a 
quark and an antiquark forms a flux tube which confines them. 
For the pentaquark there would be four such flux tubes connecting the  
quarks with the antiquark. 

\noindent
{\it Mass spectrum of $\Theta$ baryons}

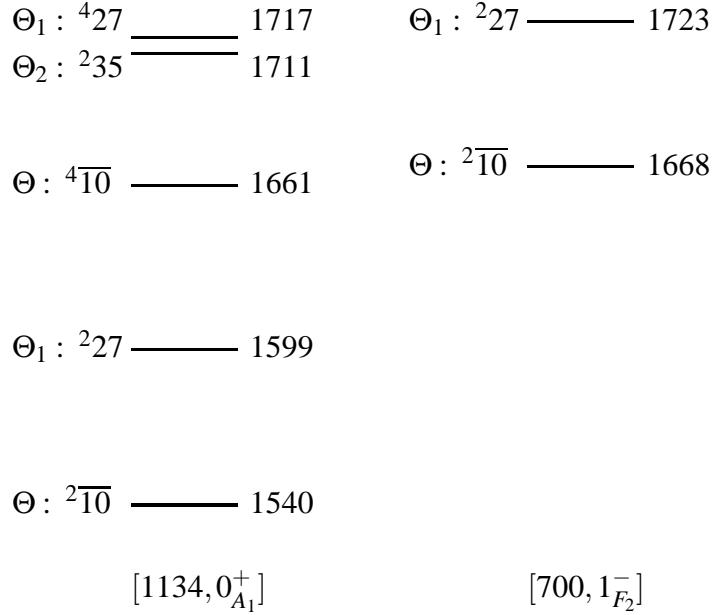
\begin{figure}[t]
\centering
\setlength{\unitlength}{1.0pt}
\begin{picture}(300,280)(0,0)

\thicklines
\put( 60, 60) {\line(1,0){ 40}}
\put( 60,181) {\line(1,0){ 40}}
\put( 60,119) {\line(1,0){ 40}}
\put( 60,237) {\line(1,0){ 40}}
\put( 60,231) {\line(1,0){ 40}}
%\put( 50, 40) {$[42111]_{F_2}$}
\put( 60, 25) {$[1134,0^+_{A_1}]$}

\put(210,188) {\line(1,0){ 40}}
\put(210,243) {\line(1,0){ 40}}
%\put(200, 40) {$[51111]_{A_1}$}
\put(210, 25) {$[700,1^-_{F_2}]$}

\put( 15, 57) {$\Theta: \; ^{2}\overline{10}$}
\put( 15,178) {$\Theta: \; ^{4}\overline{10}$}
\put( 15,116) {$\Theta_1: \; ^{2}27$}
\put( 15,240) {$\Theta_1: \; ^{4}27$}
\put( 15,222) {$\Theta_2: \; ^{2}35$}

\put(165,185) {$\Theta: \; ^{2}\overline{10}$}
\put(165,240) {$\Theta_1: \; ^{2}27$}

\put(105, 57) {1540}
\put(105,178) {1661}
\put(105,116) {1599}
\put(105,240) {1717}
\put(105,222) {1711}

\put(255,185) {1668}
\put(255,240) {1723}
\end{picture}
\vspace{15pt}
\caption{Spectrum of $\Theta$ pentaquarks. Masses are given in MeV.}
\label{theta}
\end{figure}

Hadronic spectra are characterized by the occurrence of linear Regge 
trajectories with almost identical slopes for baryons and mesons. 
Such a behavior is also expected on basis of soft QCD strings in which the 
strings elongate as they rotate. In the same spirit as in algebraic models 
of stringlike $q^3$ baryons \cite{BIL}, we use the mass-squared operator 
\ba
M^2 \;=\; M_0^2 + M_{\rm vib}^2 + M_{\rm rot}^2 + M_{\rm sf}^2 ~. 
\label{mass}
\ea
The vibrational term $M_{\rm vib}^2$ describes the vibrational spectrum 
corresponding to the normal modes of of a tetrahedral $q^4 \overline{q}$ 
configuration 
\ba
M_{\rm vib}^2 \;=\; \epsilon_1 \, \nu_1 
+ \epsilon_2 \, \left( \nu_{2a}+\nu_{2b} \right)  
+ \epsilon_3 \, \left( \nu_{3a}+\nu_{3b}+\nu_{3c} \right)  
+ \epsilon_4 \, \left( \nu_{4a}+\nu_{4b}+\nu_{4c} \right) ~. 
\label{mvib}
\ea
The rotational energies are given by a term linear in the orbital angular 
momentum $L$ which is responsable for the linear Regge trajectories in 
baryon and meson spectra 
\ba
M^2_{\rm rot} \;=\; \alpha \, L ~.
\label{mrot}
\ea
The spin-flavor part is expressed in a G\"ursey-Radicati form, 
i.e. in terms of Casimir invariants of the spin-flavor groups of 
Eq.~(\ref{sfchain})
\ba
M^2_{\rm sf} \;=\; a \, C_{2SU_{\rm sf}(6)} + b \, C_{2SU_{\rm f}(3)} 
+ c \, C_{2SU_{\rm s}(2)} + d \, C_{1U_{\rm Y}(1)} 
+ e \, C_{1U_{\rm Y}(1)}^2 + f \, C_{2SU_{\rm I}(2)} ~.
\label{msf}
\ea
The coefficients $\alpha$, $a$, $b$, $c$, $d$, $e$ and $f$ are taken 
from a previous study of the nonstrange and strange baryon resonances 
\cite{BIL}, and the constant $M_0^2$ is determined by identifying the 
ground state exotic pentaquark with the recently observed 
$\Theta(1540)$ resonance. Since the lowest orbital states with $L^p=0^+$ 
and $1^-$ are interpreted as rotational states, for these excitations there 
is no contribution from the vibrational terms $\epsilon_1$, $\epsilon_2$, 
$\epsilon_3$ and $\epsilon_4$. The results for the lowest $\Theta$ pentaquarks 
(with strangeness $+1$) are shown in Fig.~\ref{theta}.

The lowest pentaquark belongs to the flavor antidecuplet with spin $s=1/2$ 
and isospin $I=0$, in agreement with the available experimental information 
which indicates that the $\Theta(1540)$ is an isosinglet. In the present 
calculation, the ground state pentaquark belongs to the $[42111]_{F_2}$ 
spin-flavor multiplet, indicated in Fig.~\ref{theta} by its dimension 1134, 
and an orbital excitation $0^+$ with $A_1$ symmetry. Therefore, the ground 
state has angular momentum and parity $J^p=1/2^-$, in agreement with recent 
work on QCD sum rules \cite{sumrule} and lattice QCD \cite{lattice}, 
but contrary to the chiral soliton model \cite{soliton}, various cluster 
models \cite{cluster} and a lattice calculation \cite{chiu} that predict a 
ground state with positive parity. The first excited state at 1599 MeV is 
an isospin triplet $\Theta_1$ state with strangeness $S=+1$ of the 27-plet 
with the same value of angular momentum and parity $J^p=1/2^-$. The lowest 
pentaquark state with positive parity occurs at 1668 MeV and belongs to the 
$[51111]_{A_1}$ spin-flavor multiplet (with dimension 700) and an orbital 
excitation $1^-$ with $F_2$ symmetry. 
In the absence of a spin-orbit coupling, in this 
case we have a doublet with angular momentum and parity $J^p=1/2^+$, $3/2^+$. 

There is some preliminary evidence from the CLAS Collaboration for the 
existence of two peaks in the $nK^+$ invariant mass distribution at 1523 
and 1573 MeV \cite{battaglieri}. The mass difference between these two 
peaks is very close to the mass difference in the stringlike model between 
the ground state pentaquark at 1540 MeV (fitted) and the first excited 
state $\Theta_1$ at 1599 MeV. 

\noindent
{\it Magnetic moments}

The magnetic moment of a multiquark system is given by the 
sum of the magnetic moments of its constituent parts 
\ba
\vec{\mu} \;=\; \vec{\mu}_{\rm spin} + \vec{\mu}_{\rm orb} \;=\; 
\sum_i \mu_i (2\vec{s}_{i} + \vec{\ell}_{i}) ~, 
\ea
where the quark magnetic moments $\mu_u$, $\mu_d$ and $\mu_s$ are determined 
from the proton, neutron and $\Lambda$ magnetic moments and 
satisfy $\mu_q=-\mu_{\bar{q}}$. 

The $SU_{\rm sf}(6)$ wave function of the ground state pentaquark 
has the general 
structure 
\ba
\psi_{A_2} \;=\; \left[ \psi^{\rm c}_{F_1} \times \left[ \psi^{\rm o}_{A_1} 
\times \psi^{\rm sf}_{F_2} \right]_{F_2} \right]_{A_2} ~. 
\ea
Since the ground state orbital wave function has $L^p=0^+$, the 
magnetic moment only depends on the spin part. For the $\Theta^+$ 
exotic state we obtain 
\ba
\mu_{\Theta^+} \;=\; (2\mu_u + 2\mu_d + \mu_s)/3 
\;=\; 0.382 \; \mu_N ~,
\label{mmneg}
\ea
in agreement with the result obtained \cite{Liu} for the MIT bag model. 
These results for the magnetic moments are independent of the 
orbital wave functions, and are valid for any quark model in which the 
eigenstates have good $SU_{\rm sf}(6)$ spin-flavor symmetry. 

\section{Summary and conclusions}

In this contribution, we have discussed a stringlike model of pentaquarks, 
in which the four quarks are located at the corners of an equilateral 
tetrahedron with the antiquark in its center. Geometrically this is the  
most stable equilibrium configuration. The ground state pentaquark belongs 
to the flavor antidecuplet, has angular momentum and parity $J^p=1/2^-$ and, 
in comparison with the proton, has a small magnetic moment. The width is 
expected to be narrow due to a large suppression in the spatial overlap 
between the pentaquark and its decay products \cite{song}.

The first report of the discovery of the pentaquark has triggered an enormous 
amount of experimental and theoretical studies of the properties of exotic 
baryons. Nevertheless, there still exist many doubts and questions about the 
existence of this state, since in addition to various confirmations there are 
also several experiments in which no signal has been observed. Hence, it is 
of the utmost importance to understand the origin between these apparently 
contradictory results, and to have irrefutable proof for or against its 
existence \cite{zhaoclose}.  
If confirmed, the measurement of the quantum numbers of the 
$\Theta(1540)$, especially the angular momentum and parity, and the excited 
pentaquark states, may help to distinguish between different models and to 
gain more insight into the relevant degrees of freedom and the underlying 
dynamics that determines the properties of exotic baryons.  

\section*{Acknowledgements}

It is a great pleasure to dedicate this article to the 60th anniversary 
of Stuart Pittel in acknowledgement of many years of collaboration and 
friendship. This work is supported in part by a grant from CONACyT, M\'exico.

\end{document}